\documentstyle[12pt]{article}
\input{epsf}
\begin{document}
\title {K-Matrix Analysis of the (${IJ}^{PC}=00^{++}$) Amplitude
in the Mass Region up to 1550 MeV.}
\author{V.V.Anisovich and A.V.Sarantsev\\
Petersburg Nuclear Physics Institute\\
Gatchina, St.Petersburg 188350, Russia\\
e-mail:anisovic@lnpi.spb.su\\
       vsv@hep486.pnpi.spb.ru}
\date{}
\maketitle
\newcommand{\be}{\begin{eqnarray}}
\newcommand{\ee}{\end{eqnarray}}
\newcommand{\inli}{\int\limits}

\begin{abstract}

 K-matrix analysis of the $00^{++}$ wave is performed
in the channels $\pi\pi,~K\bar K,~\eta\eta$ and $4\pi$ in the mass
region up to 1550 MeV. The fit is based on the following data: $p\bar p
~(at~rest)\to \pi^0\pi^0\pi^0,~\pi^0\pi^0\eta,~\pi^0\eta\eta$
\cite{crybar,spanier}, $\pi N\to \pi\pi N$ \cite{cern,gams},
$\pi N\to K\bar K N$ \cite{bnl}
and the inelastic cross section of the $\pi\pi$ interaction \cite{alston}.
Simultaneous
analysis of these data confirms the existence of the scalar
resonances:  $f_0(980),~f_0(1300)$ and $f_0(1500)$, the
poles of the amplitude being at the following complex masses (in
MeV): $(1008\pm 10)- i(43\pm 5)$, $(1290\pm 25)-i(120\pm 15)$,
and $(1497\pm 6)-i(61\pm 5)$.  The fourth pole has sunk deeply into
the complex plane: $(1430 \pm 150) - i(600\pm 100)$. Positions of the
K-matrix poles (which are  referred to the masses of bare states) are at
$750\pm 120$ MeV, $1240\pm 30$ MeV, $1280\pm 30$ MeV and $1615\pm 40$ MeV.
Coupling constants of the K-matrix poles to the $\pi\pi,~\eta\eta$ and
$K\bar K$ channels are found that allow us to analyze the quark
and gluonic content of  bare
states. It is shown that $f_0^{bare}(1240)$ and
$f_0^{bare}(1615)$ (which are strongly related to $f_0(1500)$)
can be considered as good candidates for  scalar glueball.

\end{abstract}

\newpage

 Identification of  scalar resonances in the mass region
800--2000 MeV and their classification in
 the $q\bar q$-nonets is the necessary step in a  search for
the lowest scalar glueball.  The couplings of
these states to the pseudoscalar mesons ($\pi\pi$, $K\bar K$,
$\eta\eta$, $\eta\eta'$, $\eta'\eta'$) are the glueball signature:
 the gluonic nature of the resonance reveals itself in
the relations for  branching ratios into these channels
\cite{close,vvan}.

Glueball spectra obtained in the QCD-inspired models
\cite{jaffe,isgur} as well
as in the framework of the lattice QCD calculations
\cite{ukqcd,weing} are backing the
expectation that the lowest scalar glueball is located in the mass region
1000--2000 MeV. However, the predictions for glueball mass can not be
taken literally. The present lattice QCD calculations of glueball mass do
not take into account  quark degrees of freedom properly, while
according to the rules of the $1/N$-expansion \cite{thooft}
the glueball can mix with
$q\bar q$-components without  suppression. A mass shift of the order
of 100--300 MeV can  easily be caused by an admixture of
$q\bar q$-components.

Scalar resonances in the region 1000--1500 MeV have been investigated in a
number of papers \cite{crybar,gams,absz,bsz} where
it was shown that the data in this mass region
can be described by the three scalar resonances: $f_0(980)$,
$f_0(1300)$ and $f_0(1500)$. However, in these papers the K-matrix
formalism has been used for the energy region below 1200 MeV, and the
resonances $f_0(1300)$ and $f_0(1500)$ have been studied in the framework of
the T-matrix technique.  The T-matrix technique which is very effective
in the determination of the poles of the physical amplitude
cannot be applied to define the initial "bare" poles
(related to the quark-antiquark or gluon-gluon interactions) because
of resonances overlapping.
The proper method to investigate the structure of the "bare"
states is the K-matrix approach or the dispersion relation $N/D$-method,
as  was repeatedly emphasized in recent times (see, for example, ref.
\cite{mpt}).

In ref. \cite{aps} the resonance
$f_0(980)$ have been analyzed in the framework of a  two-channel
($\pi\pi$ and $K\bar K$) K-matrix formalism. The expansion of the
K-matrix analysis into the region of higher mass requires
accounting for other open channels, namely, $\eta\eta$ and $4\pi$.
In the present paper we perform the four-channel K-matrix analysis
of the following experimental data: meson spectra in the reactions
$p\bar p~(at~rest)\to \pi^0\pi^0\pi^0,~
\eta\pi^0\pi^0,~\eta\eta\pi^0$ obtained by Crystal
Barrel collaboration \cite{crybar,spanier}, two-pion spectra in the
reactions $\pi^-p\to \pi^-\pi^+ n$  (CERN-M\"unich collaboration
\cite{cern}) and $\pi^-p\to\pi^0\pi^0 n$ (GAMS \cite{gams}), partial cross
 section $\pi\pi\to K\bar K$ in the $00^{++}$-wave \cite{bnl} and
the inelastic cross section for the $\pi\pi$ interaction \cite{alston};  the
channels $\pi\pi,~K\bar K,~\eta\eta$ and $4\pi$ are taken into
account.

Our analysis confirms the
existence of three scalar resonances
$f_0(980)$,  $f_0(1300)$ and $f_0(1500)$;
the fourth pole of the scattering amplitude may correspond to a
broad resonance at 1300--1600 MeV, $f_0(1300/1600)$. But the
K-matrix solution is not unique. We have found two sets of solutions which
describe well the data and give approximately the same positions for the
poles of the physical amplitude and for the K-matrix poles.
Coupling constants of
the K-matrix poles to $\pi\pi,~K\bar K$ and $\eta\eta$ also coincide
reasonably. But the solutions differ in
the K-matrix background terms.

In the leading terms of the $1/N$ expansion, quark combinatorics predict
definite ratios for coupling constants of $q\bar q$ mesons decaying
into two pseudoscalar mesons: these ratios depend on the flavour content
of the decaying meson. Based on this, we show that the only lowest
bare state has  large $s\bar s$ component, approximately the same as in
the $\eta$-meson. The others have a small
admixture of $s\bar s$. Two of them, $f_0^{bare}(1240)$ and
$f_0^{bare}(1615)$, are
good candidates for a glueball: their coupling constants $f_0^{bare}
\to \pi\pi, K\bar K,\eta\eta$ coincide with predictions for the glueball
decay.

{\bf 1) K-matrix technique.}
We use  standard K-matrix technique for the description of the meson
scattering amplitudes in the $00^{++}$-state:
\be
\hat A=\hat K(\hat I-i\hat\rho\hat K)^{-1},
\ee
where $K_{ab}$ is a $4\times 4$ matrix ($a,b$=1,2,3,4) with the following
notations for meson states: 1=$\pi\pi$, 2=$K\bar K$, 3=$\eta\eta$
and 4=$\pi\pi\pi\pi$. The phase space matrix is a diagonal one:
\be
\rho_{ab}=\delta_{ab}\rho_a\;,\qquad
\rho_{a}=\sqrt{1-\frac{4m_a^2}{s}},\qquad \mbox{ for }\quad a=1,2,3,
\ee
$m_a$ being meson masses ($m_1=m_\pi$, $m_2=m_K$ and
$m_3=m_\eta$) and $s$ is the invariant energy squared.
The phase space factor for the four-pion state is defined
at $s<1$ GeV$^2$ either as two-$\rho$-meson phase
space or as  two-$\sigma$-meson phase space and as  unity at
$s > 1$ GeV$^2$. Thus, for two $\rho$-mesons, $\rho_4(s)$ is
expressed by:
\be
\rho_4(s) =\left\{ \begin{array}{cl}
\rho_0\int\frac{ds_{1}}{\pi}\int
\frac{ds_{2}}{\pi}\frac{M^2\Gamma(s_{1})\Gamma(s_{2})
\sqrt{(s+s_{1}-s_{2})^2-4ss_{1}}}
{s[(M^2-s_{1})^2+M^2\Gamma^2(s_{1})]
[(M^2-s_{2})^2+M^2\Gamma^2(s_{2})]} & s<1\;GeV^2\\
1 & s>1\;GeV^2\;\;\;.  \end{array} \right.
\ee
Here $s_{1}$ and $s_{2}$ are two-pion energies squared,
$M$ is the $\rho$-meson  mass and
$\Gamma(s)$ is its energy-dependent width:
\be
\Gamma(s)=\gamma\rho_1^3(s)
\ee
The factor $\rho_0$ provides the continuity of
$\rho_4(s)$ at $s=1$ GeV$^2$. For the $\sigma \sigma $
intermediate state (the $\sigma$-meson is the $00^{++}$-state)
the phase space factor at $s<1$ GeV$^2$  has  more cumbersome
structure than that of eq.(3); however the results are practically
the same  as for the $\rho \rho $ phase space.

For  $K_{ab}$ we use a parametrization similar to that of
ref. \cite{aps}:
\be
K_{ab}(s)=\left ( \sum_\alpha \frac{g^{(\alpha)}_a g^{(\alpha)}_b}
{M^2_\alpha-s}+f_{ab}\frac{1+s_0}{s+s_0} \right )\;
\frac{s-m_\pi^2/2}{s},
\ee
where $g^{(\alpha)}_a$ are coupling constants of the bare state $\alpha$
to meson channels; parameters $f_{ab}$ and $s_0$
describe the smooth part of the K-matrix elements ($s_0>1.5$ GeV).

Production processes are considered using the technique of
initial-state-vector           of ref. \cite{aitch}.
Amplitudes of  $\pi\pi,~K\bar K$ and $\eta\eta$ production
in  $\pi N$ collisions due to $t$-channel pion exchange can be written as
follows:
\be
A_{\pi N\to N a}=g(\bar\Psi_N\gamma_5\Psi_N)F_N(t)D(t)
A_{\pi\pi(t)\to a},
\ee
where $a=\pi\pi,~K\bar K$ or $\eta\eta$; $F_N(t)$ is the nucleon form
factor, $D(t)$ is the pion propagator (or  reggeized pion
propagator, for more detail see ref. \cite{aps}).
 $A_{\pi\pi(t)\to a}$ is
the transition amplitude which depends on the virtuality of the
$t$-channel pion. In the initial-state-vector approach, this
amplitude is of  the  form:
\be
A_{\pi\pi(t)\to a}=P_b(1-i\hat\rho \hat K)^{-1}_{ba}.
\ee
Here  $\vec P$ depends on the virtuality of the
initial pion. This vector corresponds to the components of the K-matrix
with  modified initial-state blocks:
\be
P_b=\left ( \sum_\alpha \frac{\tilde g^{(\alpha)}(t)
g^{(\alpha)}_b} {M^2_\alpha-s}+\tilde f_{b}(t)\;
\frac{1+s_0}{s+s_0} \right )\;\frac{s-m_\pi^2/2}{s}.
\ee
In the limit $t\to m_\pi^2$, the expression (7) turns into the
scattering amplitude, so:
\be
\tilde g^{(\alpha)}(t\to m_\pi^2)=g_1^{(\alpha)}\qquad\qquad
\tilde f_{b}(t\to m_\pi^2)=f_{1b}.
\ee

The part of the amplitudes $p\bar p~(at~rest)\to \pi^0\pi^0\pi^0,
~\pi^0\eta\eta,~\pi^0\pi^0\eta$
which corresponds to the two-meson production in $00^{++}$ states
can be written in the following
way \cite{absz1}:
\be
A_{p\bar p\to mesons}=A_1(s_{23})+A_2(s_{13})+A_3(s_{12}).
\ee
The amplitude $A_k(s_{ij})$ stands for any interaction of
particles in intermediate states but the last interaction is
that of  particles $i$ and $j$ in the state $00^{++}$,
the particle $k$ being a spectator. For
 $\pi^0\pi^0\pi^0$ production $A_1=A_2=A_3$,
and the amplitude $A_k$ which describes the final-state interaction of
the $\pi^0\pi^0$ pair in the $00^{++}$ state is equal to:
\be
A_k(s_{ij})=P^{(\pi)}_b(s_{ij})\left(1-i\hat \rho \hat
K\right)^{-1}_{b1}.
\ee
The components of $\vec
P^{(\pi)}$, where index $(\pi)$ characterizes the spectator
particle, are as follows:
\be
P^{(\pi)}_b(s_{ij})=\left (
\sum_\alpha \frac{\Lambda_\alpha^{(\pi)} g^{(\alpha)}_b}
{M^2_\alpha-s_{ij}}+\phi^{(\pi)}_b\; \frac{1+s_0}{s_{ij}+s_0} \right
)\;\frac{s_{ij}-m_\pi^2/2}{s_{ij}}.
\ee
Parameters $\lambda_\alpha^{(\pi)}$ and $\phi_b^{(\pi)}$ may be complex
magnitudes with different phases.

The same vector $\vec P^{(\pi)}$ defines the
amplitude of the $00^{++}$ $\eta\eta$-state
in the reaction $p\bar p \to\pi^0\eta\eta$:
\be
A_1(s_{23})=P^{(\pi)}_b(s_{23})\left(
1-i\hat \rho \hat K\right)^{-1}_{b3}.
\ee
Production of the $\pi^0\pi^0$ pair in the reaction $p\bar p
\to\eta\pi^0\pi^0$  is described by  eqs. (11) and (12),
with the replacement  of indices $(\pi) \to (\eta)$.
The couplings $\Lambda_\alpha^{(\eta)}$
and $\phi_b^{(\eta)}$ represent  other sets of complex parameters.

{\bf 2) Coupling constants in the decays $f_0\to\pi\pi$, $K\bar K$,
$\eta\eta$, $\eta\eta'$, $\eta'\eta'$ in the quark combinatorics.}
In the leading terms of the $1/N$ expansion, the couplings of
 $q\bar q$-mesons and glueball decays to two
mesons are determined
by the diagrams shown in Figs. 1$a,b$. In these processes,  gluons produce
either one $q\bar q$-pair ($q\bar q$ decay, Fig. 1a) or two $q\bar q$ pairs
(glueball decay, Fig. 1b). The production of  soft $q\bar q$ pairs by
gluons goes on with the flavour symmetry violation: the direct
indication of such a violation comes from  multiparticle production
in the central hadron collisions at high energies
(see ref. \cite{ahkm} and references therein) and from
radiative $J/\Psi$
decays \cite{vvan}. In these cases the production of strange quarks is
suppressed by the same factor $\lambda$. The ratios of the production
probabilities are:
\be
u\bar u:d\bar d:s\bar s=1:1:\lambda
\ee
with $\lambda=0.4-0.5$. Extending this property to the
decays of $00^{++}$ $q\bar q$-mesons  allows us
to calculate the ratios of  coupling
constants $f_0\to\pi\pi$, $K\bar K$, $\eta\eta$, $\eta\eta'$,
$\eta'\eta'$. They are given in Table 1 for $f_0=n\bar
n\;\cos\Phi+s\bar s\;\sin\Phi$, where $n\bar n=(u\bar u+d\bar d)/\sqrt 2$.

In Table 1 we also present  the coupling constants taken
from ref. \cite{vvan} for the decay of a
glueball. One can see that the glueball decay-couplings
satisfy the same relations as $q\bar q$-meson couplings,
with the  mixing angle $\Phi=\Phi_{glueball}$ where
\be
\tan\Phi_{glueball}=\sqrt{\frac{\lambda}{2}}.
\ee
This is the result of the two-stage decay of the glueball (see Fig.1b):
the intermediate $q\bar q$-state in the glueball decay is the mixture
of  $n\bar n$ and $s\bar s$ quarks with the probabilities given by
eq.(14). With $\lambda=0.4-0.5$, eq. (15) goes to
$\Phi_{glueball}=25^o\pm 3^o$.

If $f_0$ is a mixture of the $q\bar q$-state ($q\bar q=n\bar n\cos
\phi+s\bar s\sin \phi$) and the glueball ($GG$),  $f_0=q\bar q\cos
\alpha+GG\sin \alpha$, then coupling constants for the decay of this
resonance into two $q\bar q$-mesons are the same as given in Table 1,
with the replacement $\Phi \to \Phi_{effective}$, where
\be
\tan \Phi_{effective}=\frac
{g \sqrt{2+\lambda}\sin \Phi+\tilde g\sqrt{\lambda}\tan\alpha}
{g \sqrt{2+\lambda}\cos \Phi+\tilde g\sqrt{2}\tan\alpha}.
\ee
The values $g$ and $\tilde g$ are determined in Table 1.

{\bf 3) Fit of the data.}
Fitting CERN-M\"unich data on the reaction $\pi^- p\to n\pi^+\pi^-$,
we use the values for moments $N<Y_L^M>$ provided
by this group \cite{cern}. The moments with $M=0$
and $M=1$ are related to the partial amplitudes as follows:
$$N<Y^0_L>\equiv\sum_{JJ'}
\sqrt{\frac{(2J+1)(2J'+1)}{4\pi(2L+1)}}<JJ'00|L0>\{<JJ'00|L0>
Re\rho_{00}^{JJ'}$$
$$
-[1+(-1)^{J+J'-L}]<JJ'1-1|L0>Re\rho^{JJ'}_{11}\},
$$
\be
N<Y^1_L>\equiv\sum_{JJ'}
\sqrt{\frac{(2J+1)(2J'+1)}{4\pi(2L+1)}}<JJ'00|L0><JJ'01|L1>
2Re\rho_{01}^{JJ'}
\ee
$$
\rho^{JJ'}_{00}=T^J_0T^{J'*}_0,\qquad \rho^{JJ'}_{01}=T^J_0T^{J'*}_-,
\qquad \rho^{JJ'}_{11}=2T^J_-T^{J'*}_-,
$$
$$
g^J_0=2C_N\sqrt{(\frac{s}{4}-m_\pi^2)(2J+1)}\; A_J(s),\qquad
g_-^J=\frac{ g^J_0}{C_J\sum^3_{n=0}a_n s^{n/2}}.
$$
Here $J$ and $J'$ are     angular momenta of outgoing pions and
$<JJ'mm'|LM>$ are Clebsch-Gordan coefficients. The amplitude $T^J_0$
describes the $t$-channel pion exchange and $A_J(s)$ is the partial
wave amplitude for the transition $\pi^-\pi^+(t)\to\pi^-\pi^+$.
The amplitudes with $J=0,1,2,3$ and  $I=0,1,2$ have been used in
the fit. The data of ref. \cite{cern} are selected with  small
$t$-values, in the interval
 $t<0.15$ (GeV/c)$^2$. So, when fitting these data, we neglect the
$t$-dependence in the $K$-matrix elements, supposing that it causes only
a renormalization of the coefficient $C_N$. For the
partial wave $IJ^{PC}=00^{++}$, we have used the K-matrix in the form of
eqs. (1)--(9), taking $\tilde g^{(\alpha)}$ and $\tilde f_a$
at $t=m_\pi^2$.
The treatment of other partial waves  is performed
in the same way as in ref. \cite{cern}.

The main contribution into $T_-^J$ is provided by the $t$-channel
$a_1$-exchange.
This amplitude is fitted in the same form as $T_0^J$ but with additional
$s$-dependent factor $C_J\sum a_n s^{n/2}$, where $C_J$ and $a_n$ are free
parameters.

 The BNL group \cite{bnl} has
performed a partial wave analysis of the experimental
data and extracted the         amplitude
squared $|A_{\pi\pi(t)\to K\bar K}|^2$.
The events have been selected in a very narrow $t$-interval,
$|t|<0.07$ (GeV/c)$^2$; so, as before, we calculate this
amplitude at $t=m_\pi^2$ too.

The $\pi^-p\to
\pi^0\pi^0n$  events obtained by GAMS \cite{gams} are
distributed over
six $t$-intervals (in (GeV/c)$^2$ units):  $0<-t<0.2$, $0.3<-t<1$,
$0.35<-t <1$,    $0.4<-t<1$,
$0.45<-t<1$ and $0.5<-t<1$.
The following $t$-dependence gives a good description of
GAMS data:
$$
\tilde g^{(\alpha)}(t)=g^{(\alpha)}_1+(1-\frac{t}{m_\pi^2})\,
\frac{t}{m_\pi^2}g'^{(\alpha)},\qquad
\tilde f_{a}(t)=f_{1a}+(1-\frac{t}{m_\pi^2})\,\frac{t}{m_\pi^2}f'_{a}
$$
\be
F_N(t)=\left [ \frac{\tilde\Lambda-m_\pi^2}
{\tilde\Lambda-t}\right ]^4, \qquad
\qquad D(t)=(m_\pi^2-t)^{-1}.
\ee
The event number  in each $t$-interval can
be calculated integrating the amplitude squared, $|A_{\pi N \to
N a}|^2$, over the corresponding $t$-interval.

When fitting  Crystal Barrel data \cite{crybar,spanier}, we use eqs.
(10)--(13) for the $00^{++}$ wave,  while the other partial waves are
treated in the same way as described in detail in ref. \cite{crybar}.

{\bf 4) Two K-matrix solutions.}
In the simultaneous fit of the data  [1--6],
we have obtained two groups of
K-matrix solutions. Inside each group, the solutions give
similar descriptions of the data; the variations are
mainly due to the form used for the $4\pi$ phase space,
or they are related to the parameters which affect the reactions weakly.
For example, at the moment  we cannot fix in the $4\pi$ channel
the relative probabilities   related to two $\rho$- and
two $\sigma$-meson states. Below we
discuss two solutions which treat the $4\pi$ channel in the
simplest way $(\pi\pi\pi\pi=\rho\rho$) and keep  weakly controlled
parameters equal to zero.
 Parameters for these solutions are given in Table 2.

Both solutions describe well the experimental data:
a typical description of the CERN-M\"unich data is
shown in Fig. 2,
GAMS data in Fig. 3; the $\pi\pi\to K\bar K$
  and $\pi\pi\to \eta\eta$ amplitude squared as well as the
$\pi\pi$ inelastic cross section are drawn in Fig.4.
Corresponding
 $\chi^2$ values
per point are given in the second and fourth columns of Table 3. In our
simultaneous fit, $\chi^2$ for every reaction is not worse than the $\chi^2$
obtained by various groups in the separate fits of these reactions.  For
example, in ref.  \cite{cern} (CERN-M\"unich data) $\chi^2$ per point is
2.6 (in our fit $\chi^2=1.8$); for $p\bar p$ annihilations $\chi^2(
3\pi^0) =1.8$, $\chi^2(\pi\eta\eta)=1.65$ and $\chi^2(\pi\pi\eta)=1.6$
\cite{spanier},
while in our fit the corresponding values are about 1.5, 1.5 and
1.4.  Relative branching ratio for $\pi^0\pi^0\pi^0$ and $\pi^0\eta\eta$
production in $p\bar p$ annihilation is 3.35 for both
solutions, which coincides well with the experimental magnitude
$3.1\pm 0.8$ \cite{spanier}.

 The two groups of solutions differ in sign in the K-matrix
non-pole term for the transition $\pi\pi\to K\bar K$
 and in the couplings of the first and third K-matrix poles to the $K\bar
K$ channel. This affects mainly the description of the $\pi\pi\to K\bar K$
channel. To improve the second solution for this reaction,
 the BNL experimental data \cite{bnl} should be scaled down by a
factor 1.20. As is seen from Fig.4a, there is a discrepancy between BNL
and Argonne \cite{argon} data for this channel, about a factor 1.5,
so scaling down the
of BNL data by a factor 1.20 could be quite justified. Let us note
that such a change of the scale could be also plausible for the first
solution, though a lesser effect is reqired.

All solutions give rather similar positions of singularities
of the scattering amplitudes. The positions of the poles
on the second and on the third sheets are given in
Table 4: they coincide well with
the poles obtained in the $T$-matrix approach \cite{bsz}.
The other stable characteristics are phase shift and
inelasticity in the $00^{++}\pi\pi$ scattering amplitude.
Our results are shown in Fig.5 for the phase shift and
in Fig. 6 for the inelasticity.

For the solutions presented in Table 2, the couplings
of the K-matrix poles with $\alpha =2,3,4$ to $\pi\pi$, $K\bar K$ and
$\eta\eta$ channels agree with the quark combinatoric relations given
in Table 1, with
$\lambda=0.45$ and $\sin\Theta=0.57$.  Therefore, we have also fitted
 the data implying for the couplings the constraint given by Table 1. The
 description of  data changed insignificantly;  $\chi^2$ obtained with
this constraint is given in the third and fifth columns of Table 3.
Relative branching ratios for $p\bar p\to \pi^0\pi^0\pi^0$ / $p\bar p\to
 \pi^0\eta\eta$ are 3.5 for the first solution and 3.6 for the second one.
Corresponding parameters (masses of the
K-matrix poles, $\pi\pi$-channel couplings $g_1$, $\Phi_{effective}$ and
$4\pi$-channel couplings $g_4$) are given in Table 5.

Let us emphasize that we do not include in our fit the
$\pi\pi\to \eta\eta$ data. Nevertheless
the two-bump structure of the $\pi\pi\to \eta\eta$
amplitude squared is reproduced in both our solutions without $f_0(1590)$.
This resonance has been seen in the $\eta\eta'$ spectrum              as
well \cite{g1590}, hence the
incorporation of $\eta\eta'$ channel into analysis is needed to make a
final conclusion
about the nature of this resonance; this is beyond our current
investigation.

{\bf 5) Where is the lowest scalar glueball?}
The lightest $f_0^{bare}$ state has a large
$s\bar s$ component: it is approximately the same
as in the    $\eta$-meson.  For  the states $f_0^{bare}(1240)$
and $f_0^{bare}(1615)$ we have not obtained a
satisfactory solution with a large $s\bar s$ component.
For $f_0^{bare}(1280)$ the $50\%$-admixture of the $s\bar s$-component
is permissible; this state may be the $SU(3)$-nonet partner for
$f_0^{bare}(750)$. On the contrary, an assumption that
$f_0^{bare}(1240)$ is the nonet partner of $f_0^{bare}(750)$
leads to significantly larger $\chi^2$ value.

For the other two states the following scenarios are possible:
\begin{enumerate}
\item There is no glueball among these states, and they are
radial excitations
of the nonstrange $q\bar q$-system, while the mixing angles of
$f_0^{bare}$'s
coincide with the glueball angle given by eq.(15) only accidentally.
In this case we
should expect the existence of two $s\bar s$-rich
$f_0$ resonances  in the region 1550--1850 MeV;
\item One of these states corresponds to the radial excitation of
   nonstrange quarks and another one is the lowest glueball.
For example, $f_0^{bare}(1240)$ is $2^1P_0$ $n\bar n$-state,
while $f_0^{bare}(1615)$ is a glueball or vice versa. In both
cases only one $s\bar s$-rich state exists in the region
1550-1850 MeV.  \end{enumerate}

Therefore, the crucial point for the identification of a glueball is to
determine the number of $f_0$ states with  large $s\bar s$
component in the mass  region 1550--1850 MeV.
To fulfill this task a
careful analysis of the reactions $\pi p\to \eta\eta n$,
$\pi p\to \eta\eta' n$,
$p\bar p\to \eta\eta'\pi$, $p\bar p\to K\bar K\pi$, together with
the data of refs. [1--6], should be made.

{\bf 6) Conclusion.}
A K-matrix analysis based on the experimental data of refs. [1--6]
has been
performed for the wave $00^{++}$ up to 1550 MeV. Two solutions have been
found which determine the positions of K-matrix poles as well as the
couplings of these poles with two pseudoscalar mesons and with
the $4\pi$-channel.

The necessity to use the K-matrix formalism is caused by strong resonance
overlapping which is the case for the $00^{++}$-wave. It leads to the
shift of
the poles  because of the transitions $resonance \to real~
mesons\to resonance$. To perform a reliable interpretation of
$q\bar q$- or $GG$-resonances, the influence of these shifts should be
removed. The obtained K-matrix poles,  $f_0^{bare}$'s, demonstrate that
the mass shift due to the transition $resonance \to real~
mesons\to resonance$ is rather large, about 100--200 MeV, and actually
resonances are  strong mixtures of $f_0^{bare}$'s and two/four-meson
components.

Our analysis shows that  coupling constants of the $f_0^{bare}$ states
to the two pseudoscalar-meson channels obey  the
quark combinatoric rules with good accuracy.
 These rules argue strongly in favour of
one of the bare states, $f_0^{bare}(1240)$ or $f_0^{bare}(1615)$,
being considered as a
candidate for a glueball. However, for a final conclusion, the
investigation of the $00^{++}$ wave in the region 1550--1850 MeV
is needed, especially in the channels strongly coupled to the $s\bar s$ state.

We thank D.V.Bugg, S.S.Gershtein, A.K.Likhoded, L.Montanet,
Yu.D.Prokoshkin and B.S.Zou for encouraging discussions and useful remarks.
We are grateful to Dr. W. Ochs for supplying CERN-M\"unich data in numerical
form, and also the GAMS and Crystal Barrel groups for making their new high
statistics data available.
This work was partly supported by the International Science
Foundation Grant N 10300.

\newpage
\begin{center}
{\bf
Figure Captions}
\end{center}

\begin{description}
\item [\bf Fig. 1.] Diagrams for the decay of (a) a
$q\bar q$-meson and (b) a glueball into two $q\bar q$-meson
states.
\item [\bf Fig. 2.]
 Fit of the $\pi\pi$ angular-moment
distributions   in the final state of the reaction
$\pi^-p \to n\pi^+\pi^-$ at 17.2 GeV/c \cite{cern}. The curve
corresponds to the first K-matrix solution with the resonance decay
couplings obeying the quark combinatoric constraints (Table 1).
\item [\bf Fig. 3.] Event
numbers {\it versus}  invariant mass of the $\pi\pi$-system for different
$t$-cuts in the $\pi^-p\rightarrow \pi^0\pi^0n$ reaction
\cite{gams}.  The solid curve corresponds to the first K-matrix solution
with resonance decay couplings obeying the quark combinatoric constraints
and the dashed curve to the second one.
\item [\bf Fig. 4.] a) The
$\pi\pi\to K\bar K$ amplitude squared:  data are taken from refs.
\cite{bnl} (circles) and  \cite{argon} (squares);
b) The ratios $\sigma(\pi\pi \to 4\pi)/\sigma(\pi\pi \to \pi\pi)$
\cite{alston};
c) $\pi\pi\to \eta\eta$ amplitude squared \cite{bnl}.
The style of the curves is the same as in Fig. 3.
\item [\bf Fig. 5.] The S-wave $\pi\pi$ phase shift obtained in the
simultaneous fit of the data [1--6].
\item [\bf Fig. 6.]  Inelasticity of
the $\pi\pi$ S-wave obtained in the simultaneous fit of the data [1--6].
\end{description}

\newpage

\begin{center}
Table 1\\
Coupling constants given by quark combinatorics for a $q\bar q$-meson
and glueball decaying into two pseudoscalar mesons;
 $\Phi$ is  mixing angle for $n\bar n$ and $s\bar s$ states and
$\Theta$ is the mixing angle for $\eta -\eta'$ mesons:
$\eta=n\bar n \cos\Theta-s\bar s \sin\Theta$ and
$\eta'=n\bar n \sin\Theta+s\bar s \cos\Theta$.
\vskip 0.5cm
\begin{tabular}{|l|c|c|}
\hline
Channel & $q\bar q$-meson decay & Glueball decay \\
~       &    couplings          & couplings \\
\hline
$\pi^0\pi^0$
& $g_\pi\equiv g\;\cos\Phi/\sqrt{2}$ &
$G_\pi\equiv \tilde g/\sqrt{2+\lambda}$\\
$\pi^+\pi^-$ & $g_\pi$ & $G_\pi$\\
$K^+K^-$ &
$g_\pi (\sqrt{2}\;
\tan\Phi+\sqrt{\lambda})/2 $ &
$G_\pi\sqrt{\lambda}$\\
$K^0K^0$ &
$g_\pi (\sqrt{2}\;
\tan\Phi+\sqrt{\lambda})/2 $ &
$G_\pi\sqrt{\lambda}$\\
$\eta\eta$ &
$g_\pi
\left(\cos^2\Theta+\sqrt{2\lambda}\;\tan\Phi\;\sin^2\Theta\right)$ &
$G_\pi
\left(\cos^2\Theta+\lambda\;\sin^2\Theta\right)$ \\
$\eta\eta'$ &
$g_\pi \sin\Theta\;\cos\Theta
\left(1-\sqrt{2\lambda}\;\tan\Phi\right)$ &
$G_\pi
\cos\Theta\;\sin\Theta\;(1-\lambda)$ \\
$\eta'\eta'$ &
$g_\pi\left(\sin^2\Theta+\sqrt{2\lambda}\;\tan\Phi\;
\cos^2\Theta\right)$ &
$G_\pi
\left(\sin^2\Theta+\lambda\;\cos^2\Theta\right)$ \\
\hline
\end{tabular}
\end{center}
\vskip 1.cm
\begin{center}
Table 2\\
Parameters (in GeV units) for the K-matrix solutions without
quark combinatoric
constraints on the coupling constants $g_n^{(\alpha)}$.
\vskip 0.5cm
\begin{tabular}{|l|rrrr|rrrr|}
\hline
~&\multicolumn{4}{|c|}{ I solution} &
\multicolumn{4}{c|}{ II solution} \\
\hline
~ & $\alpha=1$ &$\alpha=2$ & $\alpha=3$ & $\alpha=4$ &
$\alpha=1$ &$\alpha=2$ & $\alpha=3$ & $\alpha=4$ \\
\hline
M        & 0.651& 1.235 & 1.269 & 1.565
& 0.858 & 1.209 & 1.281 & 1.583 \\
$g_{1}^{(\alpha)} $ & 1.012 & 0.800 & 0.185 & 0.484
& 0.694 & 0.961 & 0.241 & 0.547 \\
$g_{2}^{(\alpha)} $ &-0.774 & 0.495 & 0.063 & 0.050
& 0.029 & 0.551 & 0.246 & 0.146 \\
$g_{3}^{(\alpha)} $ & 0.073 & 0.228 & 0.045 & 0.155
&-0.081 & 0.232 & 0.053 & 0.167 \\
$g_{4}^{(\alpha)} $ & 0   & 0   & 0.537 & 0.383
& 0   & 0   & 0.556 & 0.455 \\
\hline
$Re(\Lambda^{(\pi)}_\alpha)$ &-0.193 &-0.013 & 2.036 & 1
&-0.186 &-0.100 & 1.832 & 1   \\
$Im(\Lambda^{(\pi)}_\alpha)$ &-0.621 &-0.641 &-0.389 & 0
&-0.509 &-0.654 &-0.014 & 0  \\
$Re(\Lambda^{(\eta)}_\alpha)$ & 1  & 0.084 &-0.770 & 0
& 1     &-0.348 &-2.484 & 0   \\
$Im(\Lambda^{(\eta)}_\alpha)$ & 0  & 0.241 & 0.145 & 0
& 0     &-1.023 &-3.112 & 0  \\
$g'_\alpha$ & 0.024 &-0.048  & 0& 0
& 0.022  &-0.046  & 0 & 0 \\
\hline
~&a=1 & a=2 & a=3 & a=4
&a=1 & a=2 & a=3 & a=4\\
\hline
$f_{1a}$ & 0.398 &-0.607 & 0 & 0
& 0.034 & 0.286 & 0 & 0.050  \\
\hline
$Re(\phi^{(\pi)}_a)$ &-0.661 & 0.434 & 0.139 &0
&-0.551 &-0.465 & 0.194 &0\\
$Im(\phi^{(\pi)}_a)$ &-0.559 & 0.413 & 0.074 &0
&-1.102 &-0.841 &-0.040 &0\\
$Re(\phi^{(\eta)}_a)$ & 1.461 &-0.643 & 0     &0
& 2.742 & 2.719 & 0     &0\\
$Im(\phi^{(\eta)}_a)$ & 1.191 &-0.943 & 0     &0
& 4.756 & 2.311 & 0     &0\\
$f'_{a}$ & 0.076 &-0.030 & 0& 0
&0.089 &-0.074 & 0& 0\\
\hline
~&\multicolumn{2}{c}{$s_0=1.8$} &
\multicolumn{2}{c|}{$\tilde\Lambda=0.149$}
&\multicolumn{2}{|c}{$s_0=5.0$} &
\multicolumn{2}{c|}{$\tilde\Lambda=0.154$}\\
\hline
\end{tabular}
\end{center}

\newpage
\begin{center}
Table 3\\
The $\chi^2$ values for the K-matrix solutions without and with quark
combinatoric constraint.
\vskip 0.5cm
\begin{tabular}{|l|c|c||c|c|}
\hline
~ & \multicolumn{2}{|c||}{ I solution} &
\multicolumn{2}{|c|}{ II solution} \\
\cline{2-5}
~ & without quark & with $g^{(\alpha)}_n$ &
    without quark & with  $g^{(\alpha)}_n$\\
~&combinatoric &  given by &
  combinatoric &  given by \\
~& constraint & Table 1 & constraint & Table 1 \\
\hline
$p\bar p\to \pi^0\pi^0\pi^0$& 1.50 & 1.53 & 1.52 & 1.57  \\
$p\bar p\to \pi^0\eta\eta$  & 1.50 & 1.50 & 1.54 & 1.53  \\
$p\bar p\to \pi^0\pi^0\eta$ & 1.34 & 1.34 & 1.39 & 1.42  \\
$\pi^+\pi^-\to\pi^+\pi^-$   & 1.75 & 1.81 & 1.72 & 1.78  \\
$ \pi^+\pi^-\to\pi^0\pi^0$  & 1.87 & 1.97 & 1.92 & 2.08  \\
$\pi\pi\to K\bar K$         & 0.68 & 0.54 & 0.88 & 0.51  \\
$\pi\pi\to 4\pi$            & 3.60 & 3.50 & 3.53 & 3.07 \\
\hline
\end{tabular}
\end{center}
\vskip 0.5cm

\begin{center}
Table 4\\
Pole positions (in MeV) in the $00^{++}$ partial wave amplitude
(2nd sheet: under $\pi\pi$ and $4\pi$ cuts; 3rd sheet:
under $\pi\pi$, $4\pi$ and $K\bar K$ cuts; 4th sheet:
under $\pi\pi$, $4\pi$, $K\bar K$ and $\eta\eta$ cuts).
\vskip 0.5cm
\begin{tabular}{|l|l|l|l|l|l|}
\hline
\multicolumn{3}{|c|}{ I solution} &
\multicolumn{3}{c|}{ II solution} \\
\hline
2nd sheet & 3rd sheet & 4th sheet &
2nd sheet & 3rd sheet & 4th sheet \\
\hline
$1015-i46$ & $936-i238$ & ~ & $1004-i41$ & $957-i41$ & ~ \\
 ~ & ~ & $1280-i116$ & ~ & ~ & $1285-i128$ \\
 ~ & ~ & $1311-i512$ & ~ & ~ & $1490-i613$ \\
 ~ & ~ & $1501-i64 $ & ~ & ~ & $1498-i64 $ \\
\hline
\end{tabular}
\end{center}
\vskip 0.5cm
\begin{center}
Table 5\\
Masses, coupling constants (in GeV) and mixing angles of the
$f_0^{bare}$-resonances.
\vskip 0.5cm
\begin{tabular}{|l|cccc|cccc|}
\hline
~ & \multicolumn{4}{c|}{ I solution} &
\multicolumn{4}{c|}{ II solution} \\
\hline
~ & $\alpha=1$ &$\alpha=2$ & $\alpha=3$ & $\alpha=4$ &
$\alpha=1$ &$\alpha=2$ & $\alpha=3$ & $\alpha=4$ \\
\hline
~ & ~& ~& ~& ~ & ~& ~&~& \\
M                   & 0.651 & 1.254 & 1.280 & 1.601
~                   & 0.805 & 1.242 & 1.265 & 1.632 \\
$\Delta M$          &$^{+.200}_{-.020}$ & $^{+.010}_{-.035}$ &
$\pm .025$ & $^{+.050}_{-.025}$
                    &$^{+.030}_{-.160}$ & $\pm .025$ &
$^{+.025}_{-.015}$ & $^{+.025}_{-.050}$\\
$g_{1}^{(\alpha)} $ & 0.869 & 0.798 & 0.261 & 0.607
~                   & 0.563 & 0.958 & 0.414 & 0.751 \\
$\Delta g_{1}^{(\alpha)}$&$^{+.200}_{-.250}$ & $\pm .050$ &
$^{+.060}_{-.100}$ & $^{+.150}_{-.060}$ &
$^{+.130}_{-.030}$ & $^{+.030}_{-.110}$ &
$^{+.030}_{-.100}$ & $^{+.020}_{-.200}$\\
$g_{4}^{(\alpha)} $ & 0     & 0     & 0.774 & 0.494
~                   & 0     & 0     & 0.624 & 0.627 \\
$\Delta g_{4}^{(\alpha)} $ & 0 & 0 &$^{+.020}_{-.200}$ &
$^{+.150}_{-.050}$ & 0     & 0     &$^{+.150}_{-.150}$
&$^{+.050}_{-.150}$   \\
 $\Phi_\alpha (deg)       $ & $-58^{+20}_{-5}$  & $17^{+10}_{-15}$
                     & $14^{+15}_{-30}$  & $-10^{+25}_{-20}$
 ~                   & $-36^{+15}_{-25}$ & $13\pm 15$ &$16\pm 30$
                     & $-5^{+30}_{-10}$ \\
\hline
~ & \multicolumn{4}{c|}{$25^o>\Phi_3-\Phi_2>0^o$} &
\multicolumn{4}{c|}{$25^o>\Phi_3-\Phi_2>0^o$} \\
\hline
\end{tabular}
\end{center}

\newpage
\begin{center}
\epsfxsize=15cm \epsfbox{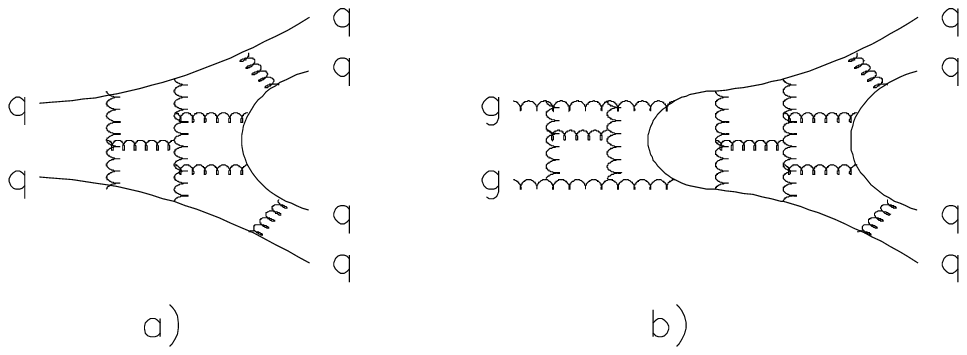}
Fig.1
\end{center}

\begin{center}
\epsfxsize=15cm \epsfbox{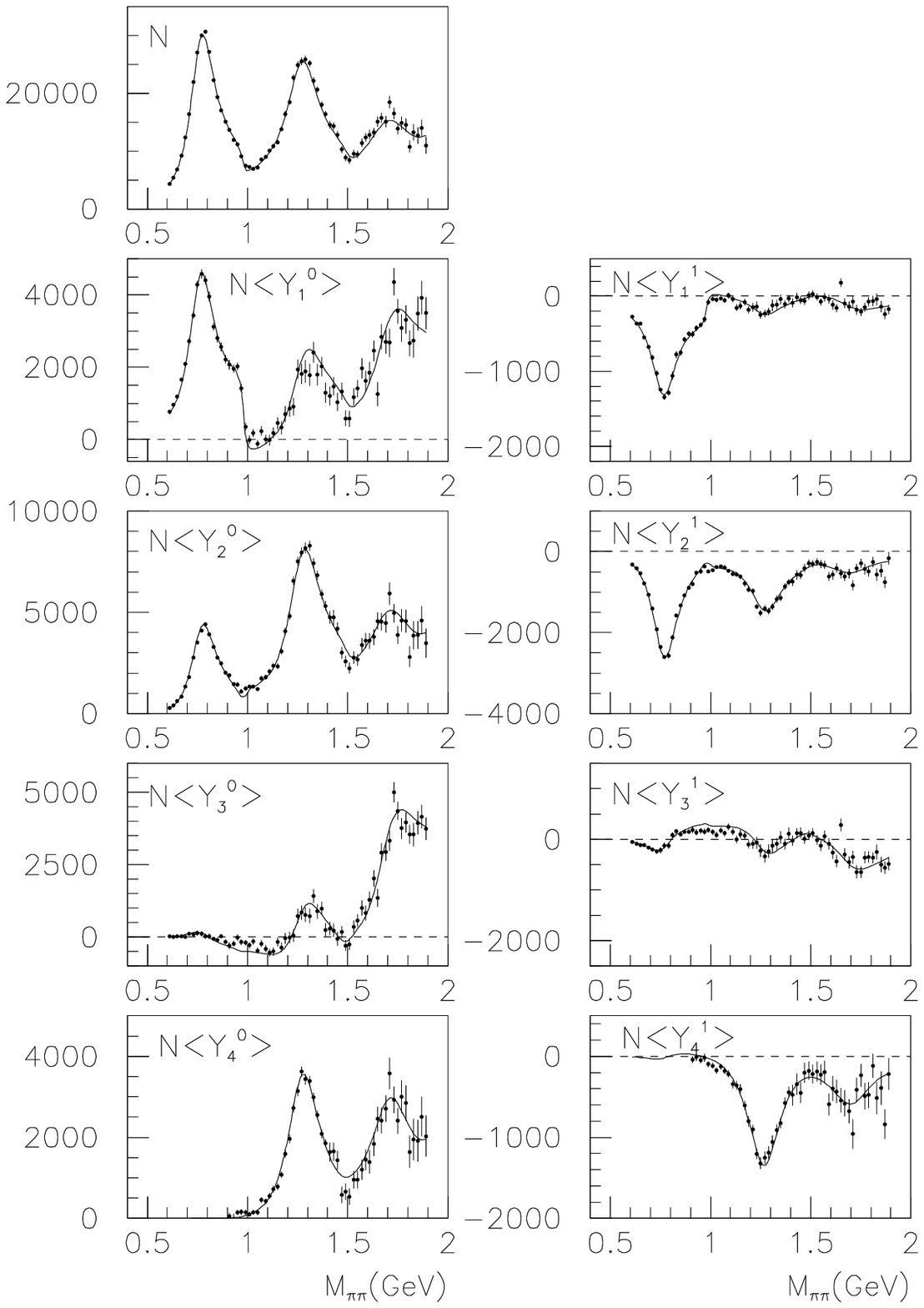}
Fig.2
\end{center}

\begin{figure}
\begin{center}
\epsfxsize=15cm \epsfbox{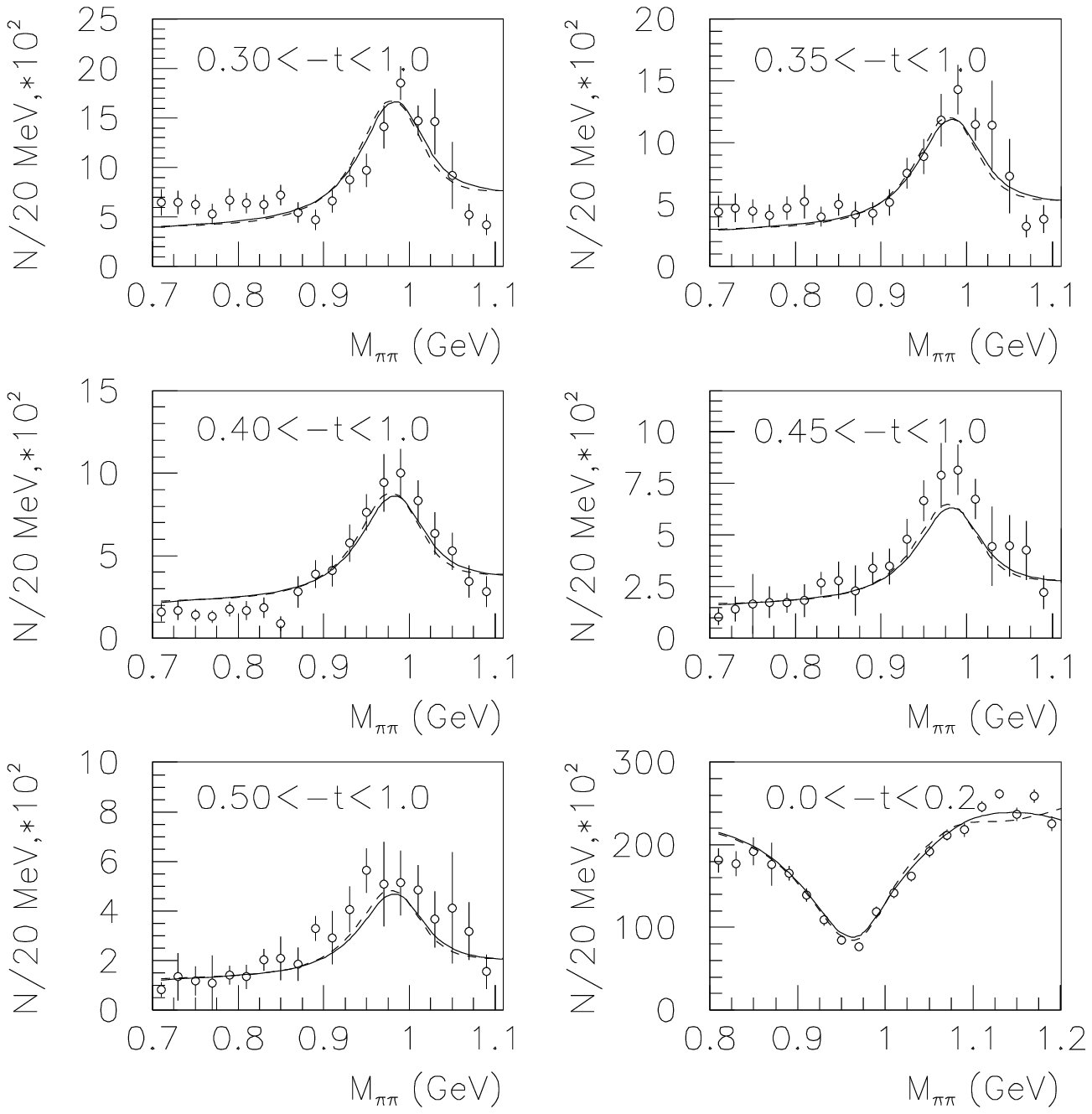}
Fig.3
\end{center}
\end{figure}

\begin{center}
\epsfxsize=10cm \epsfbox{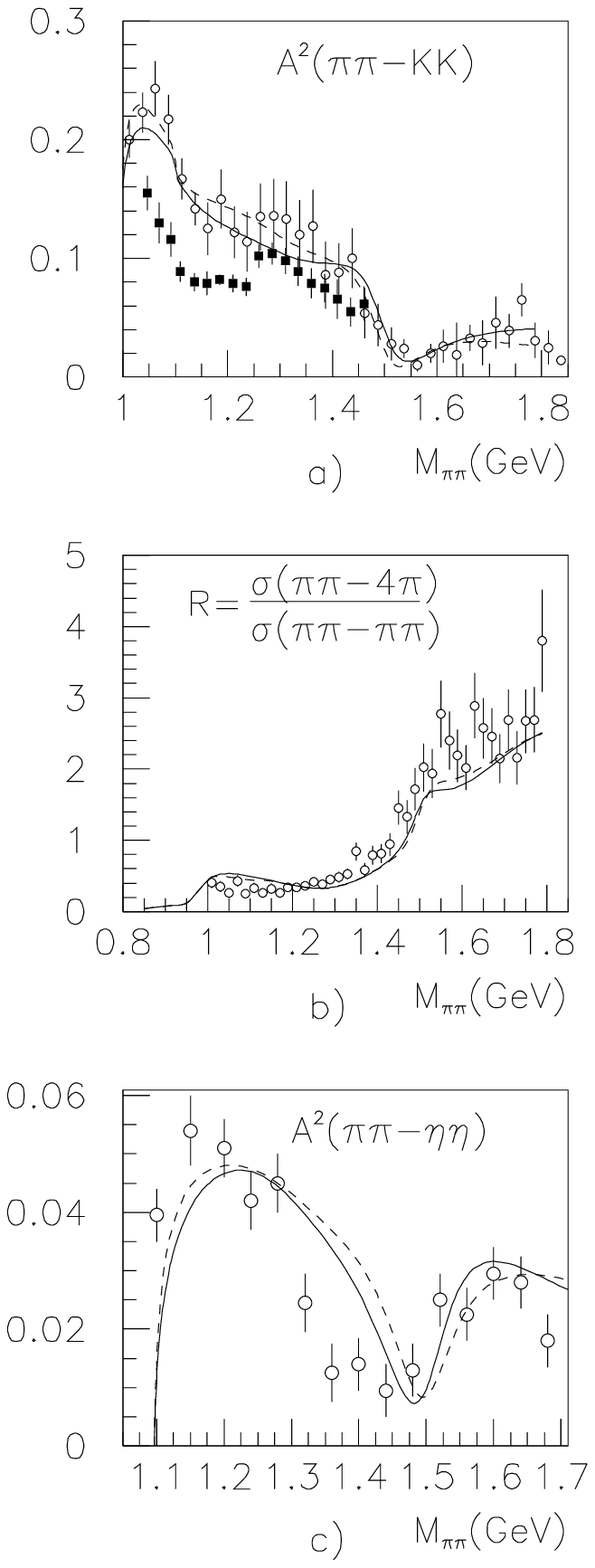}
Fig.4
\end{center}

\begin{center}
\epsfxsize=15cm \epsfbox{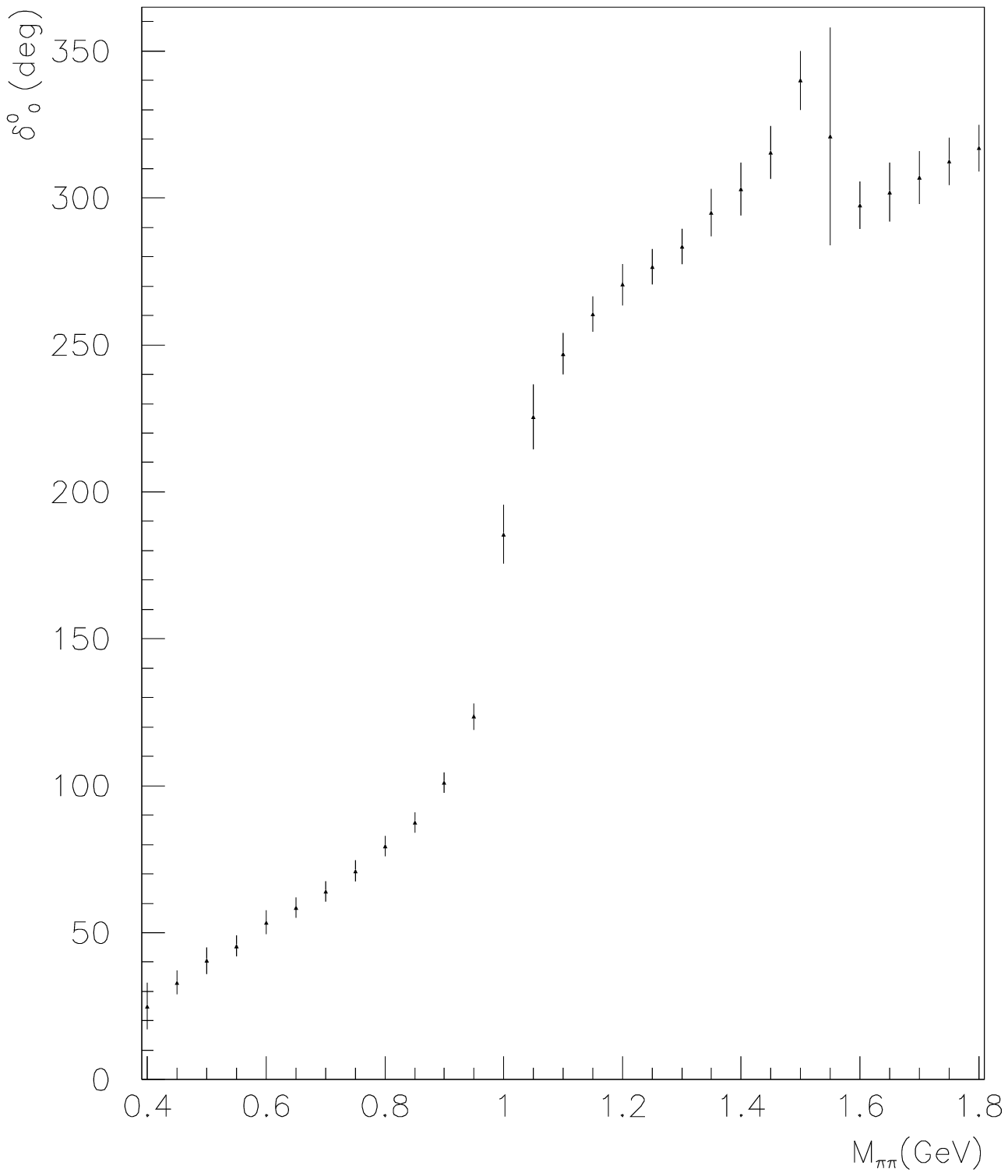}
Fig.5
\end{center}

\begin{center}
\epsfxsize=15cm \epsfbox{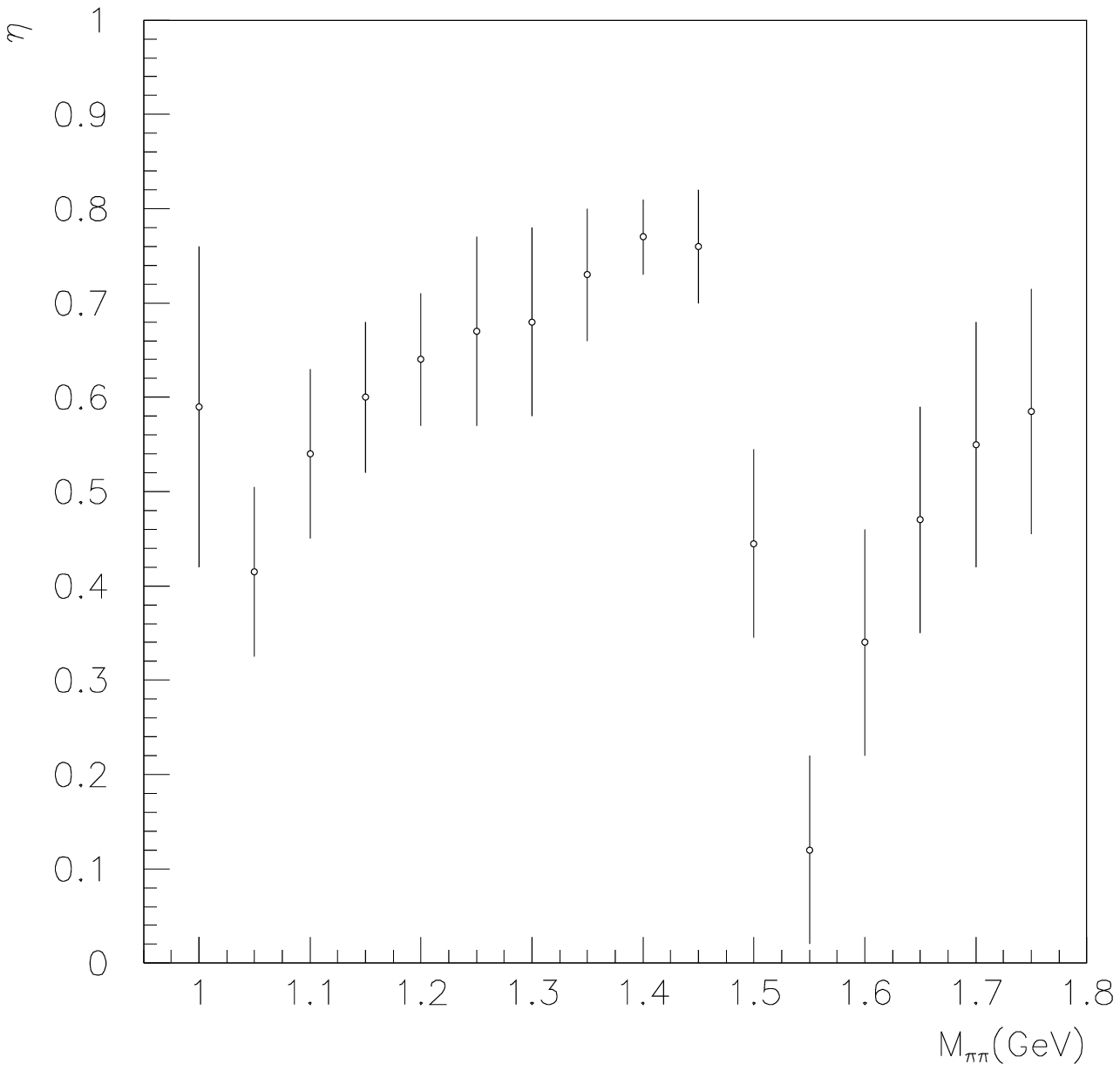}
Fig.6
\end{center}

\end{document}